# Using generative AI to investigate medical imagery models and datasets

## Authors
Oran Lang[1], Doron Yaya-Stupp[1], Ilana Traynis[2], Heather Cole-Lewis[1], Chloe R. Bennett[3], Courtney R. Lyles[1,4], Charles Lau[1], Michal Irani[5], Christopher Semturs[1], Dale R. Webster[1], Greg S. Corrado[1], Avinatan Hassidim[1], Yossi Matias[1], Yun Liu[1], Naama Hammel[1], Boris Babenko[1]

[1]Google, Mountain View, CA, USA
[2]Work done at Google via Advanced Clinical, Deerfield, IL, USA
[3]Work done at Google via Pro Unlimited, Folsom, CA, USA
[4]University of California San Francisco, Department of Medicine, San Francisco, CA USA
[5]Weizmann Institute of Science, Israel

## Abstract

**Background**
AI models have shown promise in performing many medical imaging tasks. However, our ability to explain what signals these models have learned is severely lacking. Explanations are needed in order to increase the trust of doctors in AI-based models, especially in domains where AI prediction capabilities surpass those of humans. Moreover, such explanations could enable novel scientific discovery by uncovering signals in the data that aren't yet known to experts.

**Methods**
In this paper, we present a workflow for generating hypotheses to understand which visual signals in images are correlated with a classification model's predictions for a given task. This approach leverages an automatic visual explanation algorithm followed by interdisciplinary expert review. We propose the following 4 steps: (i) Train a classifier to perform a given task to assess whether the imagery indeed contains signals relevant to the task; (ii) Train a StyleGAN-based image generator with an architecture that enables guidance by the classifier ("StylEx"); (iii) Automatically detect, extract, and visualize the top visual attributes that the classifier is sensitive towards. For visualization, we independently modify each of these attributes to generate counterfactual visualizations for a set of images (i.e. what the image would look like with the attribute increased or decreased); (iv) Formulate hypotheses for the underlying mechanisms, to stimulate future research. Specifically, present the discovered attributes and corresponding counterfactual visualizations to an interdisciplinary panel of experts so that hypotheses can account for social and structural determinants of health (e.g. whether the attributes correspond to known patho-physiological or socio-cultural phenomena, or could be novel discoveries).

**Findings**
To demonstrate the broad applicability of our approach, we present results on eight prediction tasks across three medical imaging modalities – retinal fundus photographs, external eye photographs, and chest radiographs. We showcase examples where many of the automatically-learned attributes clearly capture clinically known features (e.g., types of cataract, enlarged heart), and demonstrate automatically-learned confounders that arise from factors beyond physiological mechanisms (e.g., chest X-ray underexposure is correlated with the classifier predicting abnormality, and eye makeup is correlated with the classifier predicting low hemoglobin levels). We further show that our method reveals a number of physiologically plausible, previously-unknown attributes based on the literature (e.g., differences in the fundus associated with self-reported sex, which were previously unknown).

**Interpretation**
Our approach enables hypotheses generation via attribute visualizations and has the potential to enable researchers to better understand, improve their assessment, and extract new knowledge from AI-based models, as well as debug and design better datasets. Though not designed to infer causality, importantly, we highlight that attributes generated by our framework can capture phenomena beyond physiology or pathophysiology, reflecting the real world nature of healthcare delivery and socio-cultural factors, and hence interdisciplinary perspectives are critical in these investigations. Finally, we plan to release code to enable researchers to train their own StylEx models and analyze their predictive tasks of interest, and use the methodology presented in this paper for responsible interpretation of the revealed attributes.

**Funding**
Google

# Research in context

**Evidence before this study**

We searched Google Scholar and PubMed for published articles and book chapters available in English related to manifestation of systemic disease and demographic information in chest radiography, external eye and fundus photography and related to interdisciplinary domain expert panels for explainable AI in health, up to May 22, 2023. Our specific search terms were "cxr", "race", "ethnicity", "bone mineral density", "eyelid margin", "meibomian glands", "conjunctiva", "cataract", "diabetes", "anemia", "retinal arteries", "retinal veins", "choroid", "smoking", "cardiovascular", "sex", and "gender". Additional search terms included "interdisciplinary", "domain expert", "health", "bias", "equity", and "social determinants". Prior studies identified well known signs of or associations with systemic disease in fundus photography and chest radiography, either by human observers or AI based models. Our previous work established that AI based models could predict systemic medical conditions from external eye photographs. Although various explainability approaches have been used to interrogate medical models, there have been no studies to date which employed a generative model to identify independent visual attributes that are most relevant for AI model classification. While prior studies highlight the need for domain expertise in explainability, there is a paucity of practical methods for implementing multidisciplinary expert review, which specifically includes social scientists and sociotechnical experts who contextualize factors related to the social determinants of health and human-computer interaction.

**Added value of this study**

We developed an approach to investigate medical imaging models and datasets where a StyleGAN-based model (StylEx) is first used to generate visualizations of independent attributes which influence a medical imaging-based classifier, followed by a multidisciplinary expert review of each attribute. To demonstrate this approach, we applied it to a number of classifiers across multiple image modalities (external eye photographs, fundus images, and chest radiographs) and prediction targets ranging from demographic information, such as race and sex, to systemic conditions such as elevated blood pressure. Our expert panel review consisted of technical, medical, and socio-technical subject matter experts who critically evaluated and discussed the highlighted features and demographic or medical association. This unique approach unearthed associations that would not likely be identified from review by technical or clinical experts alone. Our study demonstrates a new approach for distinguishing the most relevant, and sometimes novel, visual attributes and holistically evaluating for bias as well as directions for future research.

**Implications of all the available evidence**

Our work shows how a generative model can be used to identify discrete medical imaging features associated with demographic information and systemic conditions. This approach allows researchers to improve their understanding of AI based models and extract new knowledge not previously identifiable by human experts. Further, we outline a multidisciplinary approach for attribute evaluation that takes into consideration the pathophysiologic and socio-cultural factors that inform healthcare delivery and thus impact our datasets, diagnosis or labels, and model development. To enable further research using this generative model, we will release the code to allow other researchers to utilize it for their modeling tasks.

# Introduction

Interest in developing Artificial Intelligence (AI) models, particularly deep learning based models for medical prediction tasks has grown considerably over the past several decades. The hope is that AI-based models can help medical providers and researchers rapidly and accurately analyze complex medical data, particularly medical imaging (e.g. radiography, endoscopy, fundus photography). Many machine learning models have performed as well as highly trained physicians[1–3], while others have reportedly achieved higher performance than their human counterparts[4–6]. While many of these prediction tasks are based on reproducing tasks that human experts are capable of, several AI models surprisingly can identify diseases and patient characteristics far beyond what the imaging modality was known to reveal[7–9]. In some cases this is due to the model picking up on confounders[10], selection/information bias in dataset selection[11,12] or structural, cultural, and historic biases reflected in data[13,14]; whereas in others the model is picking up on previously unknown manifestations of a condition[15,16].

Explainability methods help provide insights into AI models[17] and the data on which these models are trained. In the medical domain, one desirable property of an explainability method is an ability to enable a panel of experts (e.g. clinicians, statisticians, social scientists, and human factors engineers) to investigate the visual cues that the model learns to perform its task. The panel can determine whether these cues are expected given the task, or whether the cues warrant further investigation, for example, because the dataset is reflecting systematic or structural bias[18], confounding[19], or risk factors[20]. Such cases are of particular interest when developing a model which determines healthcare access, and might require steps to modify the model or the dataset on which the model was trained to ensure accuracy and limit bias. Alternatively, such a technique might lead to hypotheses for novel research directions that could improve our understanding of physiology and disease.

Among explainability methods in computer vision[21], the output of most techniques[22,23,24] is a heatmap of per-pixel importance, for instance based on the saliency of the image features. While these explainability methods can provide information about the *spatial location* (the "where") of important features, they do not typically explain higher-level features of the pixels in the highlighted region is predictive, such as texture, shape, or size (the "what"), limiting their utility in explaining possible underlying mechanisms. Recently a new line of research[25–29] showed how generative models can be used to transform images of one class into another, i.e. creating counterfactual images for these classes. While these methods show how an image changes when the class is changed, they cannot disentangle individual fine-grained attributes.

To address these challenges, in this paper we build on a StyleGAN-based approach called StylEx[30], which is able to generate visualizations of fine-grained attributes. A notable benefit of this StyleGAN-based approach beyond that of CycleGAN-based approaches is that attributes are generated using different axes of the style space. This separation enables the extraction of separate attributes that may otherwise have been merged into one, thus avoiding issues with some attributes being hidden by the most visually striking or familiar change. These attributes can potentially enable exploration of new avenues of scientific inquiry for AI models that have been found to identify surprising associations beyond what is identifiable by humans. In this study, we suggest a framework to use AI classifiers to learn new insights from medical data (Figure 1). Our approach is based on 4 fundamental steps that encompass both technical and interpretation tasks: (i) Train a classifier on a given image dataset to perform the desired machine learning task. If the overall accuracy of the trained classifier is high, then the information for visual assessment is assumed to be present in those images. (ii) Train a "StylEx" model – a StyleGAN-based image generator guided by the classifier – on this dataset of images. The loss function encourages the GAN to generate images which both resembles the original input and are correctly classified by the classifier, forcing the generator to put an emphasis on the visual attributes which are most relevant for classification. (iii) Automatically extract the top attributes which affect the classifier. (iv) Generate a visualization of these attributes for subject matter experts' evaluation, discussion, and consensus. As part of this framework, an interdisciplinary panel including both data domain experts and social scientists then interpret the attributes for consistency with prior literature and reason about whether any could be novel discoveries. Importantly, the interdisciplinary nature of the panel reduced blind spots based on individual expertise.

To demonstrate this approach, we selected three different imaging modalities and 8 different prediction tasks based on recent literature, including both "positive control' predictions and predictions that clinicians are not trained to perform using these images. The tasks were: elevated glycosylated hemoglobin (HbA1c), low hemoglobin (Hgb), and cataract presence in external eye photography; systolic hypertension, smoking status, and self-identified sex in retinal fundus photography; and abnormality and race/ethnicity in chest radiography (CXR). Generated counterfactual visualizations (i.e. what that image would look like with the attribute increased or decreased) for each attribute were reviewed by an interdisciplinary panel of machine learning engineers, social and socio-technical scientists, and modality-specific clinical specialists: a comprehensive ophthalmologist for fundus and external eye photographs, and a cardiothoracic radiologist for CXR. Based on the attribute visualizations, our specialists formed hypotheses as to what signals these attributes were reflecting. This resulted in both expected findings (i.e. visual cues that one may expect to find for a given task, which served as sound positive controls) and surprising discoveries for each imaging modality. Hypotheses generated by the interdisciplinary panel suggest future areas for research needed to better understand the observed results.

# Methods

In this section we present our end-to-end framework for automatic visual explanation of medical findings. Our method consists of 4 stages, 3 algorithmic and a final manual stage involving expert panel discussion (Figure 1).

## Stage 1: Training a classifier

In order to verify that we can indeed extract the visual information to explain the task at hand, we first train a classifier C to predict the task label. Then we test its performance on a held out set and measure its generalization capability. We apply our method on classifiers which achieve high performance (defined for the purposes of this study as being approximately above 0.8, a threshold described qualitatively as "excellent"[31]). The motivation behind this decision is that we want to ensure that the classifier has indeed learned information relevant to the task of interest, so that it is meaningful to visualize this learned information. Further details are provided in Supplementary Note 3.

## Stage 2: StylEx

We used StylEx[30] to generate counterfactual visual explanations. StylEx relies on the fact that StyleGAN2 contains a disentangled latent space called "StyleSpace"[32], which contains semantically meaningful attributes of the images. We train a StyleGAN2 model on the same dataset that is used to train the classifier, to generate realistic looking images from that domain. However, the original formulation of StyleGAN2 training does not depend on the classifier C (from Stage 1), and so it may not necessarily represent and generate subtle attributes that are important for the decision of the specific classifier C which we wish to use for explainability. Therefore, we train a StyleGAN2-like generator G to preserve the decisions of the classifier C on the training images, encouraging its StyleSpace to explicitly represent classifier-relevant attributes.

This is achieved by training the StyleGAN2 generator G with three additional changes: (i) Instead of using a mapping network from a noise vector to the latent space W, we replaced it with an encoder E, trained together with the GAN with a reconstruction-loss. This forces the generated output image to be visually similar to the input, and allows us to apply the generator on specific input images to generate counterfactuals. (ii) Together with the reconstruction loss, the classifier-output similarity loss encourages that both the generated image itself and the neuronal activations through the classification model for the generated image resemble that of the original input images. Thus, visual features important for the classifier C (such as those used to identify medical conditions) should be included in the generated image, and our expert panel's (detailed next) goals are to interpret for the presence of known and possibly novel visual features. (iii) In addition, we use a conditional version of StyleGAN2 architecture, which concatenates the classifier output of the original image to the latent vector W. This ensures that the StyleSpace vector, which is a linear mapping of this concatenated vector, uses the information from the classifier directly. To automatically find attributes which are correlated with the labels and some pre-existing set of confounding variables, we used a slightly modified method to the original StylEx work, as described in Supplementary Note 1.

## Stage 3: Automatic Attribute selection

After training the StylEx model, we search the StyleSpace of the trained generator G for attributes that affect the output of the classifier C. To do so for a given image, we manipulate each StyleSpace coordinate of G by increasing and decreasing it with a fixed factor times the standard deviation of this feature, and measure its effect on the classification prediction (henceforth "CP" for brevity), and select the top k attributes that maximize the change in CP, in a way which resembles Individual Conditional Expectation (ICE)[33]. This provides the top k image-specific attributes. We can then regenerate the input image while modifying one attribute at a time, visually displaying the meaning of each individual attribute (i.e. a counterfactual visualization). The class-specific top k attributes are obtained by repeating this process for a large number of images per class and measuring the percentage of images where CP was changed by more than some threshold (0.15), which was the same for all classifiers. To avoid outlier attributes which transformed the image out of the domain distribution, we filtered out attributes which changed the classifier prediction in opposite directions.

## Stage 4: Evaluation by interdisciplinary human experts

In our final stage, the interdisciplinary expert panel review allows us to critically assess the model findings and identify areas of bias as well as directions for future research. We visually display the top k attributes to a panel of human experts by generating counterfactual images where each attribute individually is increased or decreased for a set of example images (Figure 1). For each domain, we consult with clinical, socio-technical, social science specialists in that domain, and machine learning engineers to inspect the attributes (see Supplementary Table 3). For each attribute our clinical specialists hypothesize what visual signal is captured by the attribute and the full expert panel hypothesizes possible interpretations for why the signal is useful to the classifier. The panel discusses all attributes, with specific focus being paid to attributes the panel has identified as potentially being representative of bias due to systematic or structural factors present in the dataset or confounding not accounted for in the model. Finally, the full panel reviews and identifies a list of research and validation hypotheses. These research and validation hypotheses can be tested either by comparing them to known phenomena from the literature, or via further studies to prove or disprove them. Without these kinds of interdisciplinary review and discussion, certain social or environmental factors may be assumed to be or perpetuated as biological/physiological phenomena (see Supplementary Note 5).

For interpretation and hypothesis generation, we draw from the socio-ecological theory[34] and social and structural determinants of health framework[35]. These models argue that health is not simply biologically determined, but rather it is shaped by myriad social, political and environmental factors across the lifecourse. These models are particularly useful for the interpretation of socially-constructed attributes such as race. However, much social science research has illustrated the ways in which social, political, and environmental exposures are physically embodied and can change our biology[36] and so we apply our theoretical framework to all attributes. Additionally, research in the human factors engineering and human-computer interactions fields have demonstrated the myriad human and environmental effects, such as human error, maintenance of the technology, and environmental conditions (lighting, temperature, storage[37–39]) the accuracy and validity of the results produced by the technology. Using this information as our theoretical underpinning, prompts to the expert panel (Supplementary Note 6) are created to aid in hypothesis generation and group discussion of interpretations. As such, interpretations may consist of both biologically and socially constructed criteria. This includes physiological conditions related to the prediction task at hand, specific ways the imaging device and participants interact, representativeness of the study population, social and structural determinants of health, or a combination of them all.

## Datasets and prediction tasks

We tested our method on three different input imaging modalities, and a variety of tasks. For each modality, we used datasets from prior research studies and retrained a classifier using the methods used in these works with minor modifications (more details are included in Table 1 and Supplementary Note 2). These classifiers are oftentimes trained in a multi-task fashion, such that they can predict many different targets (i.e. there is a common "backbone", and a different "head" for each prediction task). For the purposes of our work, we selected a subset of tasks that had a sufficiently high AUC to ensure that the classifier C learned a meaningfully strong signal. In all cases, we trained the StylEx model on the same training data that was used to train the classifier models. Specifically, for fundus photos, we follow the approach reported in Poplin et al.[9], using the UK Biobank dataset; the prediction tasks in this domain include self-reported sex, systolic blood pressure (SBP)≥140, and smoking status. For external eye photos, we use the model reported in Babenko et al.[40], which was trained on the EyePACs/LACDHS dataset; the prediction tasks in this domain include presence of cataracts, HbA1c≥9, Hgb<11. Finally, for CXR we explore two prediction tasks: CXR abnormality and race. For the former we follow the approach described in Nabulsi et al.[41], using the IND1 and CXR-14 datasets. For the latter we trained a model similar to that described in Gichoya et al[13]. For uniformity across tasks and setups, all tasks were framed as binary classification problems.

Overall, most datasets were collected in a healthcare setting, and thus skewed older. The datasets also had unequal proportions of sex, and had racial/ethnic distributions which may not precisely represent the countries from which data were collected. While all the datasets used here have been previously described in the literature, because nuances of the populations and data collection processes affect the interpretation of the attributes, we include a brief summary of these datasets in Supplementary Note 2.

## Ethics

Given this study was retrospective and used de-identified datasets, the need for further review was waived by the Advarra Institutional Review Board (IRB).

## Role of Funder

Google was involved in the design and conduct of the study; management, analysis, and interpretation of the data; preparation, review, and approval of the manuscript; and decision to submit the manuscript for publication.

# Results

Across the imaging domains and prediction tasks explored in this paper, our method unveiled a number of attributes, leading to several novel areas for additional scientific inquiry according to interdisciplinary expert consensus. Table 2 lists a sample of the discovered attributes, and the full list can be found in the Supplementary Material. For each attribute, we provide a "visual explanation", by showing two generated versions of an image with increased / decreased values of the detected attribute (animated images showing these changes are attached as Supplementary Material), along with the panel's description and interpretation of the attribute, and directions for future exploration. In addition to their primary task, we trained our classifier to also predict additional variables (such as sex and age if not already being predicted), which enabled us to explore potential confounding effects (see Methods and Supp. Note 1). These results illustrate the breadth and diversity of data types and attributes for which our method is applicable and the variety of research hypotheses inspired by it. We review the highlights of these findings for each modality below, noting that while our results suggest testable hypotheses generated by the interdisciplinary panel, they do not establish causality (further discussed in the Discussion).

## Fundus photo

For the fundus photo modality we explored three tasks: predicting smoking status, systolic blood pressure over 140 mmHg, and sex[9]. For the first two tasks, our method produced attributes related to retinal vasculature. Specifically, retinal vein dilation was correlated with a higher CP of being a smoker, whereas arteriolar narrowing was correlated with higher CP of elevated systolic blood pressure. Both of these associations have been reported in the literature[42].

For the sex prediction task, our method found an attribute that associates greater choroidal vasculature visibility with increased CP of male sex. This is the converse of what previous research in this area has suggested[43–46]. Our panel hypothesized that dataset-specific factors, such as that related to distribution of axial length/myopia or fundus pigmentation within male and female populations, may drive the differences identified by the model, and further investigation is warranted. Further research into the association between sex, myopia[47], and fundus pigmentation in this dataset may help understand this link.

## External eye photo

In the external eye domain[40,48], we explored three tasks: cataract presence, low Hgb (< 11 g/dL), poor glycemic control (HbA1c ≥ 9%). The cataract presence task acted as a positive control since unlike the other tasks explored, cataract presence is feasible for a human to perform (and indeed is part of the reason these images were captured in this dataset). Indeed, our method found two attributes that are known to clinicians[49]: presence of cortical cataract spokes, extending from peripheral to central lens, and dimmer red reflex were both associated with higher CP of cataract presence.

For the low Hgb task, our method found an attribute showing that decreased conjunctival vessel prominence was correlated with increased CP of low hemoglobin, which our panel noted as being consistent with known biological phenomenon[50]. Another attribute appeared to show that increased eyeliner (a cosmetic used around the base of the eyelashes) thickness and density was associated with higher CP of low Hgb. Abnormally low hemoglobin levels is known to be a common condition in females of reproductive age[51]. Indeed, the CP of female sex was also correlated with this eyeliner attribute. Hence our panel hypothesized this may represent the presence of confounding picked up by the classifier that would need to be mitigated before considering the use of such a model.

Finally, for the poor glycemic control task, we report on two noteworthy attributes. The first shows that increased corneal arcus thickness from the limbus is associated with lower CP of elevated HbA1c. Generally, corneal arcus is associated with increased age[52], and indeed we found this correlation between this attribute and CP of age above 60. The direction of the correlation with HbA1c and the attribute is therefore surprising since HbA1c tends to *increase* with age[53]. However, closer examination of the dataset led our panel to hypothesize that this may be due to survivorship bias – the dataset comes from a diabetic retinopathy screening program, where the sickest patients (i.e. with higher HbA1c and referrable disease) are referred out of the program, and hence older patients who are still in the program tend to be healthier. The second attribute shows that an increase in eyelid margin pallor was associated with higher CP of elevated HbA1c. Our panel hypothesized that this may be related to subtle signs of changes in meibomian glands since meibomian gland disease has been reported to be more prevalent in individuals with diabetes[54,55]. However, further investigation is required to determine the specific mechanisms involved, and these findings would not signal the use for this visual feature in isolation as a screening tool.

## Chest X-ray

In the CXR domain we explored two tasks: predicting abnormality and predicting race. For predicting abnormality, StylEx highlighted three attributes, the first of which was left ventricular enlargement being correlated with higher CP of abnormality. Our panel categorized this as a known clinical phenomenon as this type of enlargement occurs in the setting of congestive heart failure, ischemic heart disease, or hypertension[56]. The second attribute showed that mediastinal widening or prominent aortic knob was associated with increasing CP of abnormality, which was also categorized as a known phenomenon since an enlarged or tortuous thoracic aorta is a common cause of mediastinal widening or prominent aortic knob in elderly individuals, particularly those with atherosclerosis[57]. The final attribute showed that more apparent over-exposure of images (i.e. darker images) were correlated with higher CP of abnormality. Our panel hypothesized that this correlation may be due to the differences between portable, or anteroposterior (AP), imagery, which tends to vary more in quality and exposure and is often taken in-patient for sicker patients, versus standard posteroanterior (PA) imagery[58]. Other hypotheses and factors to consider or warrant further investigation for this attribute are listed in Table 2.

For the task of predicting Black race from CXR, we report on three attributes. The first attribute shows that increased skeletal conspicuity is associated with increased CP of Black race. Increased skeletal conspicuity may be a radiographic indicator of increased bone mineral density, which studies have reported varies among racial/ethnic groups[59]. However, our panel noted that due to the historical scientific racism still present in medical literature and research today[60,61] and the widely accepted understanding that race is socially, rather than biologically, constructed[62–64], we cannot conclude these differences are related to biological differences. Differences in bone density may be the result of environmental exposures or structural artifacts[65,66] that are not measured in our dataset, suggesting further investigation is required. The next two attributes our panel categorized as being likely confounders. One showed that increased upper lung volume was associated with decreased CP of Black race. Increased upper lung volume is a sign of chronic obstructive pulmonary disease (COPD), and although some studies have suggested a lower prevalence of COPD among Black individuals[67,68], other studies have reported decreased COPD screening in Black populations[69]. Our panel noted that there are no known racial differences in lung volume – this variation tends to be related to socio-behavioral and environmental factors[70,71]. The final attribute showed that a more superior clavicle position relative to the lung apices was correlated with decreased CP of Black race. Our panel hypothesized that this correlation could be due to associations of race distribution and AP versus PA chest radiography. A more superiorly positioned clavicle is associated with portable AP CXRs, which are usually obtained at the hospital bedside with a patient in a recumbent position. As hospitalization rates are noted to be higher in non-hispanic Black patients[72,73], a higher frequency of AP CXR imaging was expected in Black patients, relative to other groups. As a more superior clavicle position is associated with AP imaging, a higher CP of Black race would have been expected for this feature, but was actually counter to our results. This unexpected observation could have been due to the unique characteristics of the training dataset which had a greater relative proportion of White patients to Black patients for AP images as compared to for PA images. Alternatively, thoracic kyphosis (excessive curvature of the thoracic spine often encountered in the setting of low bone mineral density) can result in more inferior, rather than more superior, clavicle position on CXR. As low bone mineral density has a weaker association with Black race than White race, one would also have expected that superior clavicle position would be associated with a higher CP of Black race - which was the opposite of what was observed. Further investigation of the dataset and model are needed to identify a complete explanation of this attribute.

# Discussion

In this paper we showcase a new explainability framework incorporating generative AI and an expert consultation process to interpret medical imaging models and datasets. Our AI generates counterfactual images using GANs, and is able to discover and visualize attributes which affect the decision of a classifier model, and hence may represent an association with the underlying task. We demonstrated our method on multiple prediction tasks and imaging modalities. Importantly, our method goes beyond an understanding of what areas of an image (i.e., the "where") are responsible for a prediction (the goal of various saliency-based methods), and helps understand what change in those regions are associated with the predictions (i.e., the "what"), with an important interpretation step to understand or generate hypotheses of physiological, social or socio-technical mechanisms that link these changes to the prediction task (i.e. the "why"). This critical last step of hypothesis generation relies on the expertise and collaborative interpretation of an interdisciplinary expert panel of clinicians, statisticians, social scientists, and human factors engineers.

Some of our tasks can be performed by human experts looking for specific visual features. The fact that our method is able to rediscover such features serves as positive control experiments. For example, cardiomegaly is a known condition visible in (and defined by measurements of the heart and chest width in) CXRs, and is evident by the counterfactual for the abnormal CXR model showing an increase in the size of the cardiac silhouette. Likewise in external eye photographs, cortical spokes are visible hallmarks of one type of cataract, and thus its presence as an attribute helps sanity check that the original classification model had learned to recognize meaningful features that define the prediction tasks. In addition, our method also rediscovers known visual cues that are not directly used to determine the ground truth label, but are known to be associated with the underlying condition. For instance, smoking increases the diameter of the blood vessels[42], which can be seen in the counterfactual attribute showing dilated vessels, but smoking status is not defined based on blood vessels' dilation status. More examples include arteriolar narrowing for elevated systolic BP and conjunctival pallor in low Hgb, both of which are known associations clinically; in both cases the attribute can be a manifestation of but do not define the respective condition. These rediscoveries provide evidence for the potential of the proposed method for scientific research.

For some of our tasks, our interdisciplinary expert panel review concluded several associations may be artifacts of dataset demographics, human interactions with technology, or the result of social and structural determinants (i.e. access to healthcare or exposure to racism and discrimination that is linked to differential hospitalization risks and utilization). For example, our method shows that low exposure increases the probability for a classifier to classify a chest X-ray as "abnormal", which could result from the patient orientation and CXR view (PA vs. AP), which is in turn associated with patient mobility, inpatient vs. outpatient status, and other factors. Because these factors often depend on local patient demographics, these associations may be spurious. Another example is the attribute which shows that adding eyeliner increases the CP for low hemoglobin. This is likely a result of social and cultural association between makeup usage and low hemoglobin levels with the same demographic factors of sex and age. Another example of social or structural determinants is the prediction of race from decreased skeletal lucency and increased conspicuity of ribs, scapulae, humeri, or thoracic vertebral bodies: simply put, bone mineral density as a predictor of race. Biological and even lifestyle factors cannot fully explain this association.[74] Further exploration is warranted regarding the effect of factors such as age distribution, environmental exposures, or nutrition. Additionally, because race is socially, rather than biologically, defined and often serves as a proxy for racism[75] and there is greater genetic variation within racial groups than between them[76], we feel it important to remind the reader that associations with race do not represent a biological phenomena. Uncovering attributes that indicate unwanted bias in the data is important in practice because such bias should be mitigated before a model is deployed. A practitioner may want to re-train the classifier with a different dataset or setup (e.g. in the case of the normal/abnormal CXR model, we could train separate models for AP and PA), or employ training methods or auxiliary data from public health datasets such as state-wide surveillance datasets, longitudinal national health surveys, and national cohort studies including factors at the individual, societal, political, and geographical levels to to add context and reduce bias[77,78].

The understanding that machine learning models can extract unwanted signals has implications for modeling and dataset design and transparency[79]. For example, researchers may desire updating model cards[80] or other transparency artifacts to ensure awareness of possible unexpected associations leveraged by the model. Moreover, this awareness could enable targeted improvements, whether by improving or modifying the training processes using custom losses or augmentations, or modifying dataset sampling during training to remove these associations[81–83]. Similarly, this knowledge could more broadly improve awareness of hidden associations that exist in both retrospective data and prospective data collection. For example, standardizing prospective data collection protocols to remove makeup, masks, jewelry, accessories, and more could help, though such efforts would need to be balanced with whether the target population during usage may not be keen on removing these. For retrospective datasets where protocols cannot be changed, more complete documentation of the data collecting environment and protocol could help researchers be more aware of potential associations.

An exciting product of our approach are attributes that might inspire hypotheses for previously unknown correlations. Such hypotheses have the potential to pave the way for novel scientific understanding, similar to the way AI techniques have been applied in non-medical domains[25,84,85]. For example, we found that increasing eyelid margin pallor was associated with increasing probability of elevated HbA1c. A possible explanation for that attribute is that it is a subtle manifestation of meibomian gland disease, which is more severe in patients with diabetes or elevated blood sugars, though the mechanism is not well understood[54,55]. Another attribute discovered by our method is that retinochoroidal pigmentation is associated with higher predicted likelihood for female sex, which adds to the evidence that sex differences have been observed in macular and peripapillary choroidal thicknesses[45,46,86]. These attribute examples are consistent with such connections, and could inspire research to better explore the underlying physiological and pathophysiological mechanisms.

It is important to note that our method was not designed to assess causality, and as such, the core assumptions of consistency, exchangeability, and positivity do not hold[87]. Prediction models are distinct from causal models in several ways, most notably in the considerations of temporality of the explanatory variables[88]. Without close attention to whether the effect truly occurred after the cause, assessing the relationship of "cause" and "effect" is impossible. However, by explicitly incorporating this understanding, interdisciplinary panel discussions helped reduce blind spots and aided in identification of both potential hypotheses (see previous paragraph) and potential confounding factors and as a result suggested opportunities for model improvement. Furthermore, it is worth noting that the workflow we present in this paper is aimed at aiding practitioners gain a better understanding of models and datasets, rather than a tool that can be employed in clinical practice (i.e. this approach as-is may not consistently aid in understanding why a model produces a particular output for a given input).

This new explainability framework contains several other notable limitations. Firstly, after attributes are extracted, their interpretation is not straightforward and requires close collaboration between the machine learning researchers and expert panel to generate hypotheses, conduct literature reviews, and even manually engineer features to validate the hypotheses. Unsurprisingly, in contrast to some of the more visually striking examples provided here, we have also found cases where the experts cannot readily interpret the counterfactuals described by our method. In other words, while the smooth animations generated by this approach render the (often subtle) change in the image visible and detectable, experts are unable to clearly describe or rationalize the change, whether from a pathophysiological perspective or otherwise. Depending on the magnitude of the change in the latent space applied to generate the counterfactual, it is also possible for the generator to produce unrealistic-looking images. Secondly, our method requires training a StyleGAN model, which works well on structured, closed domains such as images taken using standardized protocols and backgrounds (CXR, fundus photographs, external eye photographs), but can struggle in general open-ended domains such as images with a wide variety of backgrounds, poses, and distances (e.g., clinical photographs in dermatology where the background and camera orientation and lighting can differ substantially across cases). Training StyleGAN models is also both computationally and data intensive at present, narrowing the number of projects where this approach is feasible. Additionally, this method is subject to selection bias, measurement error, and other forms of confounding unless explicitly accounted for. Along these lines, it is important to consider the role of social and structural determinants of health in shaping both the health outcomes captured in datasets and its representativeness. Datasets used in our analyses are not necessarily demographically representative of their geographic source populations and thus our findings about potential attributes and their associations are not always generalizable to each respective country in which data were collected. Furthermore, because social, political, and economic circumstances vary vastly between countries, findings from data collected in one country cannot be generalizable to another country (i.e. data from the U.S. may not be generalizable to India).

In summary, we have presented a StyleGAN-based method (StylEx) for generating counterfactuals to explain AI models, and demonstrated its utility across multiple imaging modalities and prediction tasks. StylEx was able to rediscover known associations, detect confounders, and generate new insights that can be tested in future studies. With this paper, we will release sample training code to aid researchers, and look forward to more novel discoveries by the community.

# Figures

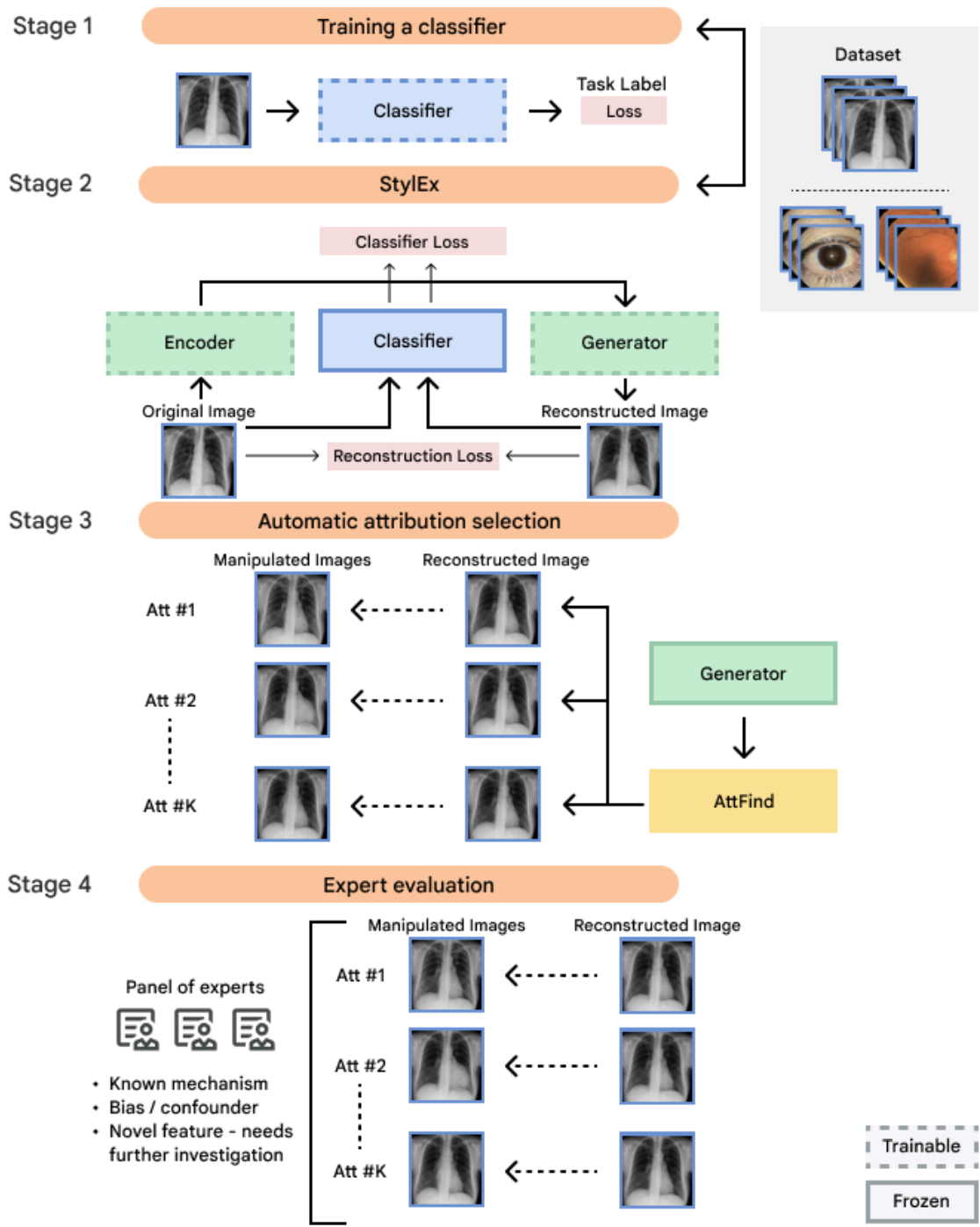

**Figure 1**

Our method consists of 4 stages: (1) train a classifier on the predictive task of interest, (2) train a StylEx model on the dataset using the classifier as guidance, (3) automatically extract the top attributes which affect the classifier, and (4) generate a visualization of these attributes for experts' evaluation.

# Tables

**Table 1**
Dataset characteristics.

| Imaging modality | External eye | Fundus photograph | Chest radiograph (CXR) | |
|---|---|---|---|---|
| Dataset | EyePACS/LACDHS | UKBiobank | Apollo & CXR-14 | MIMIC III[89] |
| Geography | Los Angeles, CA, USA | United Kingdom | India / USA | Boston |
| Number of patients | 49,025 | 62,631 | 228,108 | 47,621 |
| Number of visits | 76,458 | 64,173 | 249,558 | 69,626 |
| Number of images | 151,267 | 119,092 | 282,604 | 162,639 |
| Age (median / IQR) | 57.8 / 12.5 | 59.5 / 13.0 | 48.0 / 21.0 | 63.0 / 26.0 |
| Race/ethnicity | Data available for: 42,715<br>Hispanic: 31,708 (64.7%)<br>Black: 4,504 (9.2%)<br>Asian / Pacific islander: 3,457 (7.1%)<br>White: 2,375 (4.8%)<br>Other: 610 (1.2%)<br>Native American: 61 (0.1%) | Data available for: 62,255<br>White: 57,344 (91.6%)<br>Other: 2,309 (3.7%)<br>Asian / Pacific islander: 1,914 (3.1%)<br>Black: 688 (1.1%) | N/A | Data available for: 47,621<br>White 33,552 (70.5 %)<br>Black 8,840 (18.6 %)<br>Hispanic 3,315 (7.0 %)<br>Asian 1,914 (4.0 %) |
| Self-reported sex=Male | 19,267 / 49,014 (39.3%) | 28,633 / 62,631 (45.7%) | 140,872 / 228,081 (61.8%) | 22,620 / 47,621 (47.5 %) |

| Task-specific statistics | Label | Cataract presence | HbA1c > 9 | Hgb < 11 | Sex=Male | Smoker | Systolic BP > 140 | Abnormal | Race=Black |
|---|---|---|---|---|---|---|---|---|---|
| | Train counts [positive / total (%)] | 2,798 / 62,213 (4.5%) | 5,273 / 19,324 (27.3%) | 2,001 / 26,289 (7.6%) | 25,032 / 54,771 (45.7%) | 4,769 / 55,976 (8.5%) | 23,162 / 55,935 (41.4%) | 79,444 / 245,065 (32.4%) | 7,972 / 42,849 (18.6 %) |
| | Tune counts [positive / total (%)] | 308 / 14,245 (2.2%) | 1,188 / 4,165 (28.5%) | 451 / 5,831 (7.7%) | 3,601 / 7,860 (45.8%) | 681 / 8,021 (8.5%) | 3,278 / 8,018 (40.9%) | 2,285 / 4,493 (50.9%) | 868 / 4,772 (18.2 %) |
| | AUC [% (CI)] | 87.3 (85.8-88.9) | 70.2 (69.0-71.5) | 79.1 (77.6-80.7) | 93.9 (93.6-94.2) | 70.8 (69.3-72.3) | 77.8 (77.0-78.5) | 96.9 (96.5-97.4) | 96.1 (95.6-96.7) |

**Table 2a**
A sample of attributes for the CXR domain. Please see Supplementary Tables 2a and 2b for more details.

| Image Modality | CXR | |
|---|---|---|
| Images of increased / decreased attribute magnitude | 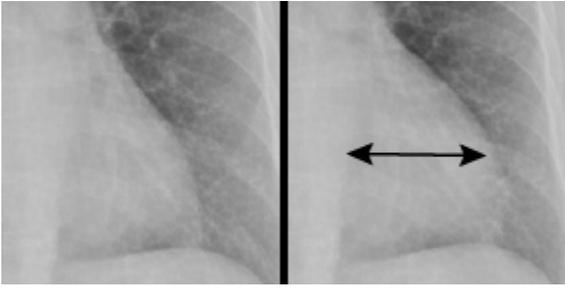 | 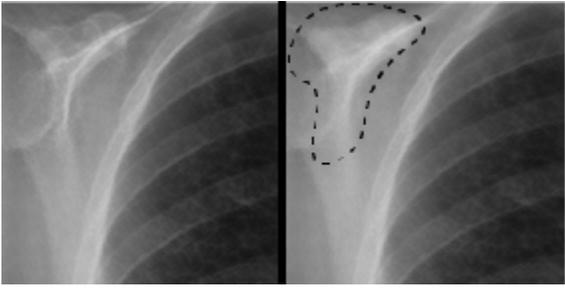 |
| Prediction task | Normal / Abnormal | Race |
| Attribute location | Left ventricle | Bones |
| Human description of the attribute ("What")" | Left ventricular enlargement, demonstrated by leftward displacement of the left heart border, with increasing CP of abnormal CXR. | Decreased skeletal lucency and increased conspicuity of ribs, scapulae, humeri, or thoracic vertebral bodies on CXR with increasing CP of Black race. |
| Consolidated panel notes | Left ventricular enlargement often occurs in the setting of congestive heart failure, ischemic heart disease, or hypertension. This attribute is unlikely to represent a regular beating of the heart because the enlargement is focal. There are severe racial disparities in heart disease and hypertension, which should be considered in the context of the dataset demographics. | Increased bone mineral density results in bones that appear more conspicuous (relative to background) on CXR. Average bone mineral density varies among racial and ethnic groups and by age. A higher average bone mineral density in Black populations may explain the association for this model; however, we cannot conclude whether the underlying cause of this is related to biological differences or environmental exposure, nutrition, or structural artifacts that are not measured. |

**Table 2b**

A sample of attributes for the Fundus domain. Please see Supplementary Table 2c for more details.

| Image Modality | Fundus photography | |
|---|---|---|
| Images of increased / decreased attribute magnitude | 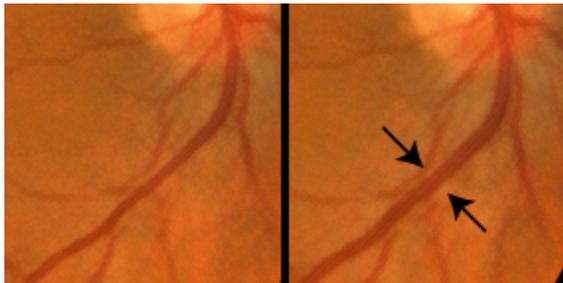 | 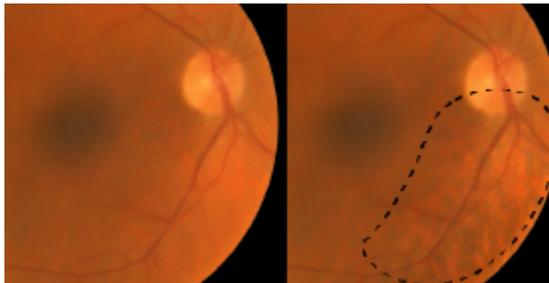 |
| Prediction task | Smoking Status (yes/no) | Sex |
| Attribute location | Retinal vessels | Choroid |
| Human description of the attribute ("What")" | Retinal vein dilation with increasing CP of being a smoker | Increase in nasal, temporal, and inferior choroidal vasculature visibility with increasing CP of male gender |
| Consolidated panel notes | Retinal venular caliber is associated with cardiovascular disease, which is strongly associated with smoking. Since smoking, cardiovascular disease, and diabetes are difficult to disentangle, it would be challenging, for example, to find datasets with smokers who do not have cardiovascular disease. Even if such datasets exist, social, cultural, and environmental factors between participants would likely vary greatly, creating additional confounders. | This attribute associates greater choroidal vasculature visibility with increased probability of male sex, which is the converse of what previous research in this area has suggested. There may be differences in the dataset, such as the distribution of myopia and fundus pigmentation within male and female populations, which may drive the differences identified by the model. |

**Table 2c**
A sample of attributes for the External Eye domain. Please see Supplementary Tables 2d and 2e for more details.

| Image Modality | External eye photograph | |
|---|---|---|
| Images of increased / decreased attribute magnitude | 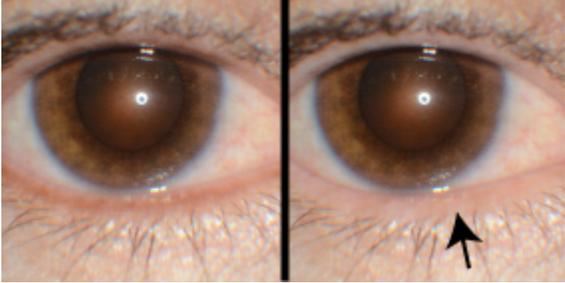 | 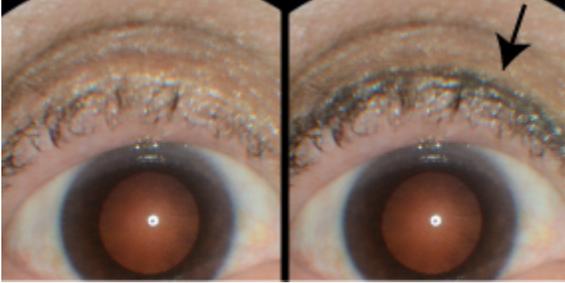 |
| Prediction task | HbA1c >= 9 | Hgb |
| Attribute location | Eyelid margin | Eyelid |
| Human description of the attribute ("What")" | Increasing eyelid margin pallor with increasing probability of elevated HbA1c | Increasing upper eyelid makeup eyeliner thickness and density with increasing CP of low Hgb |
| Consolidated panel notes | The eyelid margin attribute may be depicting pathophysiologic signs of meibomian gland disease (MGD), which is associated with diabetes. Signs and symptoms of MGD or dysfunction are noted with increased frequency in individuals with diabetes. There are several other potential confouders which should be considered, such as toxic exposures to medications or environmental irritants, social and cultural practices that are related to hygiene and use of makeup or other topicals. | The eyeliner attribute identifies a known cultural association between makeup usage and low hemoglobin levels with the same demographic factors of sex and age. Additionally, the dataset was not racially and ethnically diverse, warranting questions of whether this same association would have been observed with darker skin tones or with use of different colors of eyeliner. |

## Data sharing statement

## Declaration of Interest

## Contributors

OL, DYS, HCL, YL, NH, and BB conceived and designed the study. OL, DYS and BB developed the models and analyzed their performance, with scientific guidance from MI. HCL, CRB and CRL designed and conducted the expert panel review process. OL, IT, HCL, CRB, CL and BB participated in the panel review sessions. OL, CRB and BB drafted the manuscript with feedback from all authors. CS, DRW, GSC, MI, AH, and YM provided strategic guidance, oversight and obtained funding. OL, DYS, and BB had access to the underlying data; OL, DYS, IT, HCL, CRB, CRL, CL, YL, NH and BB had access to the generated attributes. All authors read and approved the final version of the manuscript.

## Acknowledgements

This research was conducted using the UK Biobank Resource under application number 65275. The authors thank Dr. Sreenivasa Raju Kalidindi and his team at Apollo Radiology International for their aid with the Apollo dataset, Andrew Sellergren and Zaid Nabulsi for help with CXR modeling infrastructure, Dr. Jorge Cuadros and Dr. Lauren P. Daskivich for their help with the EyePACS/LACDHS dataset, Elvia Figueroa and the LAC DHS TDRS program staff for data collection and program support, Nikhil Kookkiri and EyePACS staff for data collection and support, and Preeti Singh for support with dataset and annotation logistics. Finally, the authors would like to thank Mat

Fleck for participating the panel sessions, Avinash Varadarajan and Yossi Gandelsman for early feedback on the project, Cameron Chen, Ivor Horn, and Lily Peng for providing feedback on the manuscript, and Tiya Tiyasirichokchai for Figure 1 design. Part of the data processing team at LAC DHS was supported by grants UL1TR001855 and UL1TR000130 from the National Center for Advancing Translational Science (NCATS) of the U.S. National Institutes of Health. The content is solely the responsibility of the authors and does not necessarily represent the official views of the National Institutes of Health, or the US Government.

# References

1  Swiecicki A, Li N, O'Donnell J, *et al.* Deep learning-based algorithm for assessment of knee osteoarthritis severity in radiographs matches performance of radiologists. *Comput Biol Med* 2021; **133**: 104334.

2  Kashou AH, Ko W-Y, Attia ZI, Cohen MS, Friedman PA, Noseworthy PA. A comprehensive artificial intelligence-enabled electrocardiogram interpretation program. *Cardiovasc Digit Health J* 2020; **1**: 62–70.

3  Stidham RW, Liu W, Bishu S, *et al.* Performance of a Deep Learning Model vs Human Reviewers in Grading Endoscopic Disease Severity of Patients With Ulcerative Colitis. *JAMA Netw Open* 2019; **2**: e193963.

4  Ellertsson S, Loftsson H, Sigurdsson EL. Artificial intelligence in the GPs office: a retrospective study on diagnostic accuracy. *Scand J Prim Health Care* 2021; **39**: 448–58.

5  Ruamviboonsuk P, Krause J, Chotcomwongse P, *et al.* Deep learning versus human graders for classifying diabetic retinopathy severity in a nationwide screening program. npj Digital Medicine. 2019; **2**. DOI:10.1038/s41746-019-0099-8.

6  Qian H, Dong B, Yuan J-J, *et al.* Pre-Consultation System Based on the Artificial Intelligence Has a Better Diagnostic Performance Than the Physicians in the Outpatient Department of Pediatrics. Frontiers in Medicine. 2021; **8**. DOI:10.3389/fmed.2021.695185.

7  Xiao W, Huang X, Wang JH, *et al.* Screening and identifying hepatobiliary diseases through deep learning using ocular images: a prospective, multicentre study. *Lancet Digit Health* 2021; **3**: e88–97.

8  Ghorbani A, Ouyang D, Abid A, *et al.* Deep Learning Interpretation of Echocardiograms. DOI:10.1101/681676.

9  Poplin R, Varadarajan AV, Blumer K, *et al.* Prediction of cardiovascular risk factors from retinal fundus photographs via deep learning. *Nat Biomed Eng* 2018; **2**: 158–64.

10 Duffy G, Clarke SL, Christensen M, *et al.* Confounders mediate AI prediction of demographics in medical imaging. *npj Digital Medicine* 2022; **5**: 1–6.

11 Kaplan RM, Chambers DA, Glasgow RE. Big data and large sample size: a cautionary note on the potential for bias. *Clin Transl Sci* 2014; **7**: 342–6.

12 Rusanov A, Weiskopf NG, Wang S, Weng C. Hidden in plain sight: bias towards sick patients when sampling patients with sufficient electronic health record data for research. *BMC Med Inform Decis Mak* 2014; **14**: 51.

13 AI recognition of patient race in medical imaging: a modelling study. *The Lancet Digital Health* 2022; **4**:

e406–14.

14  Richardson R, Schultz J, Crawford K. Dirty Data, Bad Predictions: How Civil Rights Violations Impact Police Data, Predictive Policing Systems, and Justice. 2019; published online Feb 13. https://papers.ssrn.com/abstract=3333423 (accessed March 5, 2023).

15  Mitani A, Traynis I, Singh P, *et al.* Retinal fundus photographs capture hemoglobin loss after blood donation. medRxiv. 2022; : 2021.12.30.21268488.

16  L'Imperio V, Wulczyn E, Plass M, *et al.* Pathologist Validation of a Machine Learning–Derived Feature for Colon Cancer Risk Stratification. *JAMA Netw Open* 2023; **6**: e2254891–e2254891.

17  Notions of explainability and evaluation approaches for explainable artificial intelligence. *Inf Fusion* 2021; **76**: 89–106.

18  Obermeyer Z, Powers B, Vogeli C, Mullainathan S. Dissecting racial bias in an algorithm used to manage the health of populations. Science. 2019; **366**: 447–53.

19  Winkler JK, Fink C, Toberer F, *et al.* Association Between Surgical Skin Markings in Dermoscopic Images and Diagnostic Performance of a Deep Learning Convolutional Neural Network for Melanoma Recognition. *JAMA Dermatol* 2019; **155**: 1135–41.

20  Berger JS, Haskell L, Ting W, *et al.* Evaluation of machine learning methodology for the prediction of healthcare resource utilization and healthcare costs in patients with critical limb ischemia-is preventive and personalized approach on the horizon? *EPMA J* 2020; **11**: 53–64.

21  Singh A, Sengupta S, Lakshminarayanan V. Explainable Deep Learning Models in Medical Image Analysis. *J Imaging Sci Technol* 2020; **6**. DOI:10.3390/jimaging6060052.

22  Marco Tulio Ribeiro University of Washington, Seattle, WA, USA, Sameer Singh University of Washington, Seattle, WA, USA, Carlos Guestrin University of Washington, Seattle, WA, USA. 'Why Should I Trust You?' ACM Conferences. https://dl.acm.org/doi/10.1145/2939672.2939778 (accessed Nov 7, 2022).

23  Simonyan K, Vedaldi A, Zisserman A. Deep Inside Convolutional Networks: Visualising Image Classification Models and Saliency Maps. 2013; published online Dec 20. DOI:10.48550/arXiv.1312.6034.

24  Zhou B, Khosla A, Lapedriza A, Oliva A, Torralba A. Learning Deep Features for Discriminative Localization. 2015; published online Dec 14. DOI:10.48550/arXiv.1512.04150.

25  Thiagarajan JJ, Thopalli K, Rajan D, Turaga P. Training calibration-based counterfactual explainers for deep learning models in medical image analysis. *Sci Rep* 2022; **12**: 1–15.

26  Dravid A, Schiffers F, Gong B, Katsaggelos AK. medXGAN: Visual Explanations for Medical Classifiers through a Generative Latent Space. 2022 IEEE/CVF Conference on Computer Vision and Pattern Recognition Workshops (CVPRW). 2022. DOI:10.1109/cvprw56347.2022.00331.

27  Narayanaswamy A, Venugopalan S, Webster DR, *et al.* Scientific Discovery by Generating Counterfactuals Using Image Translation. *Medical Image Computing and Computer Assisted Intervention – MICCAI 2020* 2020; : 273–83.

28  DeGrave AJ, Cai ZR, Janizek JD, Daneshjou R, Lee S-I. Dissection of medical AI reasoning processes via physician and generative-AI collaboration. medRxiv. 2023; published online May 16. DOI:10.1101/2023.05.12.23289878.

29  Mertes S, Huber T, Weitz K, Heimerl A, André E. GANterfactual—Counterfactual Explanations for Medical Non-experts Using Generative Adversarial Learning. *Front Artif Intell* 2022; **5**: 825565.

30  Lang O, Gandelsman Y, Yarom M, *et al.* Explaining in Style: Training a GAN to explain a classifier in StyleSpace. 2021 IEEE/CVF International Conference on Computer Vision (ICCV). 2021. DOI:10.1109/iccv48922.2021.00073.

31  Hosmer DW Jr, Lemeshow S, Sturdivant RX. Applied Logistic Regression. John Wiley & Sons, 2013.

32  Wu Z, Lischinski D, Shechtman E. StyleSpace Analysis: Disentangled Controls for StyleGAN Image Generation. 2021 IEEE/CVF Conference on Computer Vision and Pattern Recognition (CVPR). 2021. DOI:10.1109/cvpr46437.2021.01267.

33  Molnar C. 9.1 Individual Conditional Expectation (ICE). 2023; published online Aug 21. https://christophm.github.io/interpretable-ml-book/ice.html#ice (accessed Sept 21, 2023).

34  Bronfenbrenner U. Ecological Systems Theory: In: Vasta. *Six theories of child development: revised formulations and current issues*.

35  Organization WH. A conceptual framework for action on the social determinants of health. 2010; : 76 p.

36  Krieger N. Ecosocial Theory of Disease Distribution: Embodying Societal & Ecologic Context. 2011; published online March 23. DOI:10.1093/acprof:oso/9780195383874.003.0007.

37  Zamzam AH, Abdul Wahab AK, Azizan MM, Satapathy SC, Lai KW, Hasikin K. A Systematic Review of Medical Equipment Reliability Assessment in Improving the Quality of Healthcare Services. *Front Public Health* 2021; **9**: 753951.

38  Proctor RW, Van Zandt T. Human Factors in Simple and Complex Systems. 2017.

39  McEntee M, Gafoor S. Ambient temperature variation affects radiological diagnostic performance. 2009.

40  Babenko B, Traynis I, Chen C, *et al.* Discovering novel systemic biomarkers in photos of the external eye. arXiv [eess.IV]. 2022; published online July 19. http://arxiv.org/abs/2207.08998.

41  Nabulsi Z, Sellergren A, Jamshy S, *et al.* Deep learning for distinguishing normal versus abnormal chest radiographs and generalization to two unseen diseases tuberculosis and COVID-19. *Sci Rep* 2021; **11**: 1–15.

42  Drobnjak D, Taarnhøj NC, Kessel L, Jørgensen T, Larsen M. Association Between Retinal Vessel Diameters and Cigarette Smoking. *Invest Ophthalmol Vis Sci* 2011; **52**: 2213–2213.

43  Cheong KX, Tan CS. Sex-Dependent Choroidal Thickness Differences in Healthy Adults: A Study Based on Original and Synthesized Data. Curr. Eye Res. 2019; **44**: 236.

44  Zeng J, Liu R, Zhang X-Y, *et al.* [Relationship between gender and posterior pole choroidal thickness in normal eyes]. *Zhonghua Yan Ke Za Zhi* 2012; **48**: 1093–6.

45  Shibata H, Sawada Y, Ishikawa M, Yoshitomi T, Iwase T. Peripapillary choroidal thickness assessed by spectral-domain optical coherence tomography in normal Japanese. *Jpn J Ophthalmol* 2021; **65**: 666–71.

46  Yang H, Luo H, Gardiner SK, *et al.* Factors Influencing Optical Coherence Tomography Peripapillary Choroidal Thickness: A Multicenter Study. *Invest Ophthalmol Vis Sci* 2019; **60**: 795–806.

47  Cumberland PM, Bountziouka V, Hammond CJ, Hysi PG, Rahi JS, UK Biobank Eye and Vision Consortium. Temporal trends in frequency, type and severity of myopia and associations with key environmental risk factors in the UK: Findings from the UK Biobank Study. *PLoS One* 2022; **17**: e0260993.

48  Babenko B, Mitani A, Traynis I, *et al.* Detection of signs of disease in external photographs of the eyes via deep learning. *Nat Biomed Eng* 2022; published online March 29. DOI:10.1038/s41551-022-00867-5.

49  Glasspool MG. Cataract. In: Springer D, ed. Atlas of Ophthalmology. 1982.

50  Kent AR, Elsing SH, Hebert RL. Conjunctival vasculature in the assessment of anemia. *Ophthalmology* 2000; **107**. DOI:10.1016/s0161-6420(99)00048-2.

51  Le CHH. The Prevalence of Anemia and Moderate-Severe Anemia in the US Population (NHANES 2003-2012). *PLoS One* 2016; **11**. DOI:10.1371/journal.pone.0166635.

52  Hashemi H, Malekifar P, Aghamirsalim M, Yekta A, Mahboubipour H, Khabazkhoob M. Prevalence and associated factors of corneal arcus in the geriatric population; Tehran geriatric eye study. *BMC Ophthalmol* 2022; **22**: 354.

53  Dubowitz N, Xue W, Long Q, *et al.* Aging is associated with increased HbA1c levels, independently of glucose levels and insulin resistance, and also with decreased HbA1c diagnostic specificity. *Diabet Med* 2014; **31**: 927–35.

54  Yu T, Han X-G, Gao Y, Song A-P, Dang G-F. Morphological and cytological changes of meibomian glands in patients with type 2 diabetes mellitus. *Int J Ophthalmol* 2019; **12**: 1415–9.

55  Yu T, Shi W-Y, Song A-P, Gao Y, Dang G-F, Ding G. Changes of meibomian glands in patients with type 2 diabetes mellitus. *Int J Ophthalmol* 2016; **9**: 1740–4.

56  Box L, Abbara S. Cardiac Imaging: The Requisites. Elsevier.

57  Isselbacher EM. Thoracic and abdominal aortic aneurysms. *Circulation* 2005; **111**: 816–28.

58  Mothiram U, Brennan PC, Robinson J, Lewis SJ, Moran B. Retrospective evaluation of exposure index ( EI ) values from plain radiographs reveals important considerations for quality improvement. Journal of Medical Radiation Sciences. 2013; **60**: 115–22.

59  Looker AC, Melton LJ 3rd, Harris T, Borrud L, Shepherd J, McGowan J. Age, gender, and race/ethnic differences in total body and subregional bone density. *Osteoporos Int* 2009; **20**: 1141–9.

60  Fausto-Sterling A. The bare bones of race. *Soc Stud Sci* 2008; **38**: 657–94.

61  Smedley A, Smedley BD. Race as biology is fiction, racism as a social problem is real: Anthropological and historical perspectives on the social construction of race. *Am Psychol* 2005; **60**: 16–26.

62  New AMA policies recognize race as a social, not biological, construct. American Medical Association. 2020; published online Nov 16. https://www.ama-assn.org/press-center/press-releases/new-ama-policies-recognize-race-social-not-biological-construct (accessed May 5, 2023).

63  Romualdi C, Balding D, Nasidze IS, *et al.* Patterns of Human Diversity, within and among Continents, Inferred from Biallelic DNA Polymorphisms. *Genome Res* 2002; **12**: 602–12.

64  Krieger N. Does racism harm health? Did child abuse exist before 1962? On explicit questions, critical science, and current controversies: an ecosocial perspective. *Am J Public Health* 2003; **93**: 194–9.

65  Bailey ZD, Krieger N, Agénor M, Graves J, Linos N, Bassett MT. Structural racism and health inequities in the USA: evidence and interventions. *Lancet* 2017; **389**: 1453–63.

66  Krieger N. Refiguring 'race': epidemiology, racialized biology, and biological expressions of race relations. *Int J Health Serv* 2000; **30**: 211–6.

67  Gilkes A, Ashworth M, Schofield P, *et al.* Does COPD risk vary by ethnicity? A retrospective cross-sectional study. *Int J Chron Obstruct Pulmon Dis* 2016; **11**: 739–46.

68  Martin A, Badrick E, Mathur R, Hull S. Effect of ethnicity on the prevalence, severity, and management of COPD in general practice. *Br J Gen Pract* 2012; **62**: e76–81.

69  Mamary AJ, Stewart JI, Kinney GL, *et al.* Race and Gender Disparities are Evident in COPD Underdiagnoses Across all Severities of Measured Airflow Obstruction. *Int J Chron Obstruct Pulmon Dis* 2018; **5**: 177–84.

70  Braun L. Race, ethnicity and lung function: A brief history. *Can J Respir Ther* 2015; **51**: 99–101.

71  Van Sickle D, Magzamen S, Mullahy J. Understanding socioeconomic and racial differences in adult lung function. *Am J Respir Crit Care Med* 2011; **184**: 521–7.

72  Doshi RP, Aseltine RH Jr, Sabina AB, Graham GN. Racial and Ethnic Disparities in Preventable Hospitalizations for Chronic Disease: Prevalence and Risk Factors. *J Racial Ethn Health Disparities* 2017; **4**: 1100–6.

73  Laditka JN, Laditka SB. Race, ethnicity and hospitalization for six chronic ambulatory care sensitive conditions in the USA. *Ethn Health* 2006; **11**: 247–63.

74  Ettinger B, Sidney S, Cummings SR, *et al.* Racial differences in bone density between young adult black and white subjects persist after adjustment for anthropometric, lifestyle, and biochemical differences. *J Clin Endocrinol Metab* 1997; **82**: 429–34.

75  Braveman P, Parker Dominguez T. Abandon 'Race.' Focus on Racism. *Front Public Health* 2021; **9**: 689462.

76  Rosenberg NA, Pritchard JK, Weber JL, *et al.* Genetic structure of human populations. *Science* 2002; **298**: 2381–5.

77  Beutel A, Chen J, Zhao Z, Chi EH. Data Decisions and Theoretical Implications when Adversarially Learning Fair Representations. arXiv [cs.LG]. 2017; published online July 1. http://arxiv.org/abs/1707.00075.

78  Zhang BH, Lemoine B, Mitchell M. Mitigating Unwanted Biases with Adversarial Learning. arXiv [cs.LG]. 2018; published online Jan 22. http://arxiv.org/abs/1801.07593.

79  Xu F, Uszkoreit H, Du Y, Fan W, Zhao D, Zhu J. Explainable AI: A Brief Survey on History, Research Areas, Approaches and Challenges. *Natural Language Processing and Chinese Computing* 2019; : 563–74.

80  Mitchell M, Wu S, Zaldivar A, *et al.* Model Cards for Model Reporting. DOI:10.1145/3287560.3287596.

81  Harini Suresh Massachusetts Institute of Technology, USA, John Guttag Massachusetts Institute of Technology, USA. A Framework for Understanding Sources of Harm throughout the Machine Learning Life Cycle. DOI:10.1145/3465416.3483305.

82. Yuzi He University of Southern California, Los Angeles, CA, USA, Keith Burghardt University of Southern California, Los Angeles, CA, USA, Kristina Lerman University of Southern California, Los Angeles, CA, USA. A Geometric Solution to Fair Representations. ACM Conferences. https://dl.acm.org/doi/10.1145/3375627.3375864 (accessed Sept 21, 2023).

83. Calmon F, Wei D, Vinzamuri B, Natesan Ramamurthy K, Varshney KR. Optimized Pre-Processing for Discrimination Prevention. *Adv Neural Inf Process Syst* 2017; **30**. https://proceedings.neurips.cc/paper_files/paper/2017/file/9a49a25d845a483fae4be7e341368e36-Paper.pdf (accessed Sept 21, 2023).

84. Guimerà R, Reichardt I, Aguilar-Mogas A, *et al.* A Bayesian machine scientist to aid in the solution of challenging scientific problems. *Sci Adv* 2020; **6**: eaav6971.

85. Udrescu S-M, Tegmark M. AI Feynman: A physics-inspired method for symbolic regression. *Sci Adv* 2020; **6**: eaay2631.

86. Ooto S, Hangai M, Yoshimura N. Effects of sex and age on the normal retinal and choroidal structures on optical coherence tomography. *Curr Eye Res* 2015; **40**: 213–25.

87. Hernán MA, Robins JM. Causal Inference: What If. Boca Raton: Chapman & Hall/CRC, 2020.

88. Laubach ZM, Murray EJ, Hoke KL, Safran RJ, Perng W. A biologist's guide to model selection and causal inference. *Proc Biol Sci* 2021; **288**: 20202815.

89. Website. DOI:10.13026/C2XW26.

90. EyePACS Photographer Manual. https://www.eyepacs.org/photographer/protocol.jsp#external_photos (accessed March 29, 2023).

91. Casillas A, Abhat A, Mahajan A, *et al.* Portals of Change: How Patient Portals Will Ultimately Work for Safety Net Populations. *J Med Internet Res* 2020; **22**: e16835.

92. Sudlow C, Gallacher J, Allen N, *et al.* UK Biobank: An Open Access Resource for Identifying the Causes of a Wide Range of Complex Diseases of Middle and Old Age. *PLoS Med* 2015; **12**. DOI:10.1371/journal.pmed.1001779.

93. Swanson JM. The UK Biobank and selection bias. *Lancet* 2012; **380**: 110.

94. Fry A, Littlejohns TJ, Sudlow C, *et al.* Comparison of Sociodemographic and Health-Related Characteristics of UK Biobank Participants With Those of the General Population. *Am J Epidemiol* 2017; **186**: 1026–34.

95. Majkowska A, Mittal S, Steiner DF, *et al.* Chest Radiograph Interpretation with Deep Learning Models: Assessment with Radiologist-adjudicated Reference Standards and Population-adjusted Evaluation. *Radiology* 2020; **294**: 421–31.

96. Wang X, Peng Y, Lu L, Lu Z, Bagheri M, Summers RM. ChestX-ray8: Hospital-scale Chest X-ray Database and Benchmarks on Weakly-Supervised Classification and Localization of Common Thorax Diseases. 2017; published online May 5. DOI:10.1109/CVPR.2017.369.

97. Box. https://nihcc.app.box.com/v/ChestXray-NIHCC/file/220660789610 (accessed March 29, 2023).

98. Johnson AEW, Pollard TJ, Shen L, *et al.* MIMIC-III, a freely accessible critical care database. *Sci Data* 2016; **3**: 160035.

99  Johnson AEW, Pollard TJ, Berkowitz SJ, *et al.* MIMIC-CXR, a de-identified publicly available database of chest radiographs with free-text reports. *Sci Data* 2019; **6**: 317.

100 Szegedy C, Vanhoucke V, Ioffe S, Shlens J, Wojna Z. Rethinking the Inception Architecture for Computer Vision. In: Proceedings of the IEEE Conference on Computer Vision and Pattern Recognition. 2016: 2818–26.

101 Bello I, Fedus W, Du X, *et al.* Revisiting ResNets: Improved Training and Scaling Strategies. *Adv Neural Inf Process Syst* 2021; **34**: 22614–27.

102 Sellergren AB, Chen C, Nabulsi Z, *et al.* Simplified Transfer Learning for Chest Radiography Models Using Less Data. *Radiology* 2022; **305**. DOI:10.1148/radiol.212482.

# Supplementary

## Supplementary Note 1: Classifiers with multiple predictions

One common research question of medical imaging prediction tasks is whether the prediction is using the association of the task at hand to a known confounder, such as sex or age. We can extend our method to detect such biases, by training a *multi-headed* classifier to predict the primary task and any confounder we wish to include (one head for each confounder). Then we make the following changes to our method:

In stage 1, We replace the classifier C with the multi-headed classifier.

In stage 2, The classification loss for the generator G sums the cross-entropy on all of the classifier heads. In addition, the conditional input to G is composed of all of the heads of the classifier C.

In stage 3, We select the top k attribute based on their change in classification of the original prediction task (first head of the classifier). For each of these attributes, we then measure the effect it has on the other heads of the classifier.

This modified method allows us to find attributes which are associated with both the task and the confounder.

In our models, we used the two confounders: self-reported sex and a binary partition of age (greater than 60 or not), except for (a) in the Fundus domain, self-reported sex classifier only used age as a confounder and (b) in the CXR domain, Abnormal classifier did not use any confounders.

## Supplementary Note 2: Additional dataset details

All external eye tasks were trained on data that comes from a diabetic retinopathy screening program at the Los Angeles County Department of Health Services (LACDHS), using the EyePACS system. All patients in this data have been diagnosed with diabetes. External eye photos were captured as part of the regular screening protocol[90]. The LACDHS primarily serves uninsured and low-income patients[91]. The HbA1c and Hgb values were extracted from patients' EMR records, while cataract presence was extracted from the ophthalmologist notes from the original fundus and external eye photograph review.

All fundus photo tasks were trained on the UK Biobank dataset[92], a prospective study that recruited volunteers between the ages of 40 and 69 across the UK. While the study aimed at recruiting a broad range of participants across socioeconomics and ethnicities, it has been noted that this dataset does suffer from selection bias[93,94]. The sex and smoking status labels were self-reported by the participant, and blood pressure was measured during the assessment visit.

For the CXR abnormality task we used two datasets. The first, originally described in Majkowska et al[95], consisted of imagery obtained from five regional hospital centers in India (Bangalore, Bhubaneswar, Chennai, Hyderabad, and New Delhi). Labels were generated by applying NLP techniques to the radiology reports (described in Nabulsi et al[41]). The second dataset was the publicly available CXR-14, collected by the NIH clinical center[96]. It was enriched for thoracic abnormalities[97]. Labels for this dataset were obtained via majority vote from three US board-certified radiologists[41].

Finally, for the task of predicting race/ethnicity from CXR imagery, we used the MIMIC III dataset, which was collected from critical care units of the Beth Israel Deaconess Medical Center[98]. The extraction of chest radiograph data was described in Johnson et al[99]. Race/Ethnicity data was self-reported by patients during admission. To optimize the chances that the classifier and generator could both be trained well, we filtered to ethnicities with more than 1000 patients, namely the fields: BLACK/AFRICAN AMERICAN, HISPANIC/LATINO, ASIAN, and WHITE. One of the attributes we found in this study was hypothesized to be related to the patient position during the image recording; thus we also present the breakdown of the orientation of the X-ray with respect to the patient (anteroposterior, AP, vs posterior-anterior, PA) for each race/ethnicity category (Supp. Table 1).

## Supplementary Note 3: Classifier details

**Fundus**

We followed the methods described in Poplin, et al [9], with slight modifications. Rather than training two separate models for classification and regression, we trained a single models with many classifier heads (using the standard cross-entropy loss for each), turning all continuous targets into categorical ones via the following cut-offs: age (30, 40, 50, 60, 70, 80), bmi (23.49, 25.64, 27.77, 30.73 – cutoffs chosen via percentiles on training data), self-reported sex, smoking status, ethnicity, Hgb (12.5, 11), HbA1c (5.7, 6.5), systolic BP (140), diastolic BP (90). We used an Inception V3[100] architecture, and trained using gradient descent with momentum of 0.9 a learning rate starting at 0.004 and decaying exponentially every 1000 steps by 0.99. We trained on the UK BioBank dataset, splitting it into a training and tuning set by patient (the tuning set was used for early stopping). Images were pre-processed and filtered by quality as described in the paper, and were resized to 587x587 pixels.

**External eye**

We followed the methods described in Babenko, et al [40], with few modifications: we resized images to 512x512 pixels, rather than 587x587, and used a single model rather than an ensemble. The hyperparameters we used were for the best performing model described in the paper (i.e. "Model 1").

**CXR (normal/abnormal)**

We followed the methods described in Nabulsi et al [41], training a binary classifier for normal vs abnormal chest x-ray images. We used a ResNetRS350 architecture[101] with an Adam optimizer and a learning rate of 0.0001. Input images were resized to 512x512. The classifier was trained on the "IND1" and CXR-14 datasets (a portion of the "IND1" dataset was used as a tune set for early stopping), as described in Sellergren et al [102].

**CXR (Race)**

We followed a similar method as Gichoya et al[13], with a few modifications. We trained a ResNet50 classifier on 3 categorical heads: ethnicity, age and sex. In ethnicity, we trained a 4-category head for: 'BLACK/AFRICAN AMERICAN', 'ASIAN', 'HISPANIC/LATINO', 'WHITE'; for all experiments, to be consistent with the other experiments which are binary predictions, and with prior relevant literature, we binarize the output of his classifier as 'BLACK/AFRICAN AMERICAN' versus rest. In age, we chose a cutoff of 60 to create a binary attribute. Input images were resized to 512x512. The classifier was trained for 30 epochs with an Adam optimizer and a learning rate of 0.00003.

## Supplementary Note 4: StylEx details

StylEx jointly trains an auto-encoder E, G and a discriminator D models in an adversarial way, to optimize the following target:

$$min_{\theta(E,G)}(max_{\theta(D)}(L_{adv} + \lambda_{rec}L_{rec} + \lambda_{cls}L_{cls}))$$

Where $L_{adv}$ is the standard GAN loss, $L_{rec}$ is the reconstruction loss of the auto-encoder, and $L_{cls}$ is the classifier loss:

$$L_{rec} = |X - G(E(X))| + LPIPS(X, G(E(X))$$
$$L_{cls} = KL(C(X), C(G(E(X))))$$

And $\lambda_{rec} = \lambda_{cls} = 0.1$.

All StyleGAN-v2 models used in this paper were trained with 512x512 image resolution, for 500k steps, with a learning rate of 0.002. The training time took 8 days on 4 TPUv3 chips. For the attribute selection dataset, 1000 images (500 from each class) were used to measure the attribute effects.

To filter the attributes (see Supplementary Table 4) automatically we used the following process: Out of the 5952 attributes in the StyleSpace, we first discarded the first 2048 attributes corresponding to the first two layers of StyleGAN, as we've noticed that they tend to create coarse and often unrealistic changes in the images. For each of the remaining attributes, we increased and decreased the attribute by 4 times its standard deviation, as measured on 1000 images, and quantified the percent of images for which the classifier prediction (CP) changed by more than 0.15, defined as P(att, class).

We then selected the attributes if they satisfied the following two criteria:

First, P(att, class) was bigger than X% for both classes, or P(att, class) was bigger than 2X% for at least one class. X was usually approximately 10%, and was manually tuned based on the classifier and domain such that approximately 10 attributes per classifier remained.

Second, P(att, class) changes were consistent between its 2 classes. In other words, if the probability of one class increased when we increased the attribute, the probability of the opposite class should increase when we decrease the attribute. This condition was necessary to filter attributes which resulted in out-of-domain generated images and unreliable classifier predictions.

# Supplementary Note 5: Initial interpretations of attributes prior to panel discussion

We found that individuals with different subject matter expertise looked at the attributes from different perspectives and provided attribute hypotheses not considered by other individuals. For example, for the race skeletal lucency and clavicle position attributes identified in CXR images, the research scientist wondered if anatomic variations or posture could account for this finding, while the medical expert identified variations in bone mineral density and clavicle position in PA vs AP films as plausible medical explanations. The social-technical experts added that the mechanisms behind bone mineral density differences can arise from multiple factors including variations in environmental exposure or nutrition.

Another example of unique hypotheses generation occurred during review of the smoking attribute in fundus photos. The medical expert identified retinal vein dilation in the images and noted that retinal vein dilation is a well known manifestation of hypertension and cardiovascular disease, which is linked to smoking. As such, they postulated that the attribute was identifying an association between smoking and cardiovascular disease. The social-technical experts hypothesized that the relationship between smoking and cardiovascular health may also be impacted by other social, cultural, and environmental factors.

A more subtle example arose in understanding the corneal arcus thickness feature. When first exposed to the feature, and initially blinded to the directionality of the attribute, medical experts noted that arcus tends to be associated with increasing age, which is likely associated with worse diabetic control over time (increased HbA1c). However, the directionality was opposite, and our final hypothesis of survivorship bias arose based on discussions with researchers with experience working on this dataset and careful consideration of the dataset source. Specifically, individuals with high HbA1c and hence referable diabetic eye disease were referred out of the program, and this was verified by sanity checking plots comparing HbA1c and rates of diabetic eye disease with age in this dataset.

Leveraging the perspectives of different subject matter experts is also helpful when considering hypotheses for confounders. In the external eye photos, the low hemoglobin attribute was initially described as being associated with "eye makeup" by the machine learning researchers. The medical expert hypothesizes that since eye makeup use is more common in individuals of female sex and lower hemoglobin levels are found in females as compared to males, the visualization showed makeup as a confounder for sex rather than a true biological sign of low hemoglobin levels. The broader panel noted that this particular makeup attribute was specifically changing black-colored eyeliner thickness and density. The socio-technical experts conjectured that this attribute might differ in datasets with individuals with darker skin tones or with use of different eyeliner colors and may be associated with more subtle aspects such as age groups in addition, suggesting use of more diverse datasets for future exploration.

Due to time constraints, not all of the attributes which were detected by the StylEx method were discussed by the panel. The attributes were pre-selected by one of the experts in each domain, based on factors such as clarity (whether the attribute was visually apparent), redundancy (e.g., whether several attributes seemed to be showing a similar change but in different locations), and realism (e.g., whether such a change appeared physiologically plausible). The complete list of attributes can be found in the supplementary zip file, which contains a html file with expert descriptions of each attribute.

## Supplementary Note 6: Expert panel guidelines

Prompts to the interdisciplinary panel of experts were created by the socio-technical researchers on the team and grounded in concepts from the theoretical literature on social and structural determinants of health applied to AI. These prompts were designed to facilitate the panel of experts in hypothesis generation, including highlighting the ways that society and policy (both historical and current factors), community (e.g., neighborhood and environmental factors), organizational (e.g., interpersonal interactions and data collection settings), and individual (e.g., socio-economic, behavioral, biological, and psychological) factors might contribute to observed statistical associations. For example, prompts regarding the physical structure of hospital rooms and user experience of imaging technology used for data collection were developed to aid discussion about organizational- and individual-level factors.

The motivation of this part of the study was to debug models, build trust and transparency by providing explanations, learn new insights, and to evaluate insights responsibility with consideration of the social and structural determinants of health. Panel members were provided with an historical overview of saliency-based explainability techniques, followed by a primer on the current StyleGAN-based model and top attribute extraction process. Example animated attributes were then provided for a simple prediction task of cat vs. dog (e.g., changes in ear shape and color, and the animals' head shape). Examples of attributes that were difficult to interpret directly were also highlighted, such as subtle and harder-to-interpret changes in the cat/dog images. Finally, attributes of the 3 modalities and 8 prediction tasks we were analyzing were provided one at a time, with 5 examples at a time (i.e., 5 animated images with the same attribute changing, and the changes in predictions shown as well). Context about the respective datasets that the images were drawn from (summarized in Table 1 and Supplementary Note 2) were also provided.

The panel review process started with reviews by the clinical experts to help describe what the model observed and to help orient subsequent reviewers (e.g., on the anatomy and what the image contains), and next by the socio-technical and social scientists (with access to the clinician notes). Next, all experts independently reviewed all notes before coming together for consensus discussions. During independent reviews, experts performed literature reviews in their respective areas of expertise, and surfaced relevant evidence useful for consensus discussions. During discussions, though consensus was not forced, agreement was generally reached by virtue of combining multiple perspectives and reasoning about the likelihood of each hypothesis and assumptions made. When multiple hypotheses were plausible (each with their own literature support), these nuances or alternative interpretations were noted in the results.

## Supplementary Table 1: Distribution of AP vs PA

Distribution of anteroposterior, AP, vs posterior-anterior, PA among the different race/ethnicities, with normalization by column.

|    | ASIAN          | BLACK/AFRICAN AMERICAN | HISPANIC/LATINO | WHITE           |
|----|----------------|------------------------|-----------------|-----------------|
| AP | 3,757 (72.1%)  | 16,015 (68.9%)         | 5,136 (66.1%)   | 79,752 (75.9%)  |
| PA | 1,450 (27.8%)  | 7,232 (31.1%)          | 2,625 (33.9%)   | 25,292 (24.1%)  |

# Supplementary Table 2a

## CXR attributes

| | | | |
|---|---|---|---|
| Images of increased / decreased attribute magnitude | 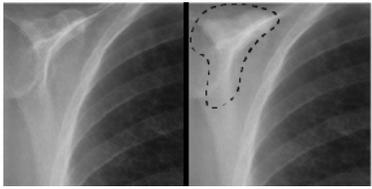 | 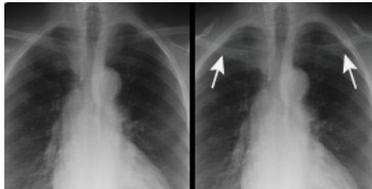 | 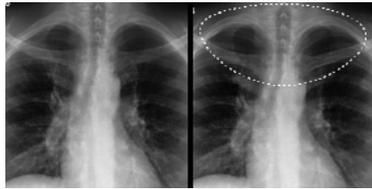 |
| Prediction Task | Race | Race | Race |
| Attribute location ("Where")" | Bones | Clavicle position | Upper lung |
| Human description of the attribute ("What")" | Decreased skeletal lucency and increased conspicuity of ribs, scapulae, humeri, or thoracic vertebral bodies on CXR with increasing CP of Black ethnicity. | More inferior clavicle position relative to the lung apices with increased CP of Black ethnicity. | Increased upper lung volume with decreased CP of Black ethnicity. |
| Consolidated Panel Notes | Increased bone mineral density results in bones that appear more conspicuous (relative to background) on CXR. Average bone mineral density varies among racial and ethnic groups and by age. A higher average bone mineral density in Black populations may explain the association for this model; however, a conclusion can't be drawn related to whether this is related to biological differences or some other environmental exposure, nutrition, or structural artifact that is not measured. Additionally, we do not know the effect the age distribution may have on these findings. | Clavicles often project superior to the lung apices on portable AP CXRs. On standard PA CXRs, clavicles generally project more inferiorly. It is possible that a difference in frequency of inpatient vs outpatient visits for black vs non-black patients affects this association. Additionally, given the racial disparities in access to care, we suggest reviewing the average number of images per patient stratified by racial group. | Larger upper lung volumes is a characteristic finding of Chronic Obstructive Pulmonary Disease (COPD). COPD is conjecture and most obvious because in normal practice this is the first consideration when seeing lungs enlarged in this way-- other considerations such as asthma and Emphysema discussed with overall understanding that CXR diagnosis depends on contextual conditions under which image was taken (e.g., ER setting would see more acute conditions vs. chronic) and that in real life, have benefit of prior images and can use that to determine acute or chronic conditions. No known racial differences in lung volume-- this variation tends to be related to socio-behavioral and environmental factors. There are racial disparities in screening and diagnosis of chronic lung diseases, such as COPD, suggesting current prevalence rates by race are inaccurate. These findings may be confounded and representative of structural bias and differential socio-behavioral factors. |
| Areas of | Would this association with CP still | Would this association with the CP | Would association with CP still be |

| Exploration from Interdisciplinary Expert Panel | stand with additional information in the model regarding unmeasured attributes such as environmental exposure, nutrition, or other structural artifact? Would this association still stand if the model were subset by age? | still exist if the model was subset by the image source (i.e. where the image was originally taken) and number of images per patient included in the data? | present if demographic (e.g. race and income) and behavioral (e.g. smoking) representation in participant admission and referral rates were different? Would we expect to find the same association if the referral rates were the same between patient racial groups? |
|---|---|---|---|

# Supplementary Table 2b

## CXR attributes

| Images of increased / decreased attribute magnitude | 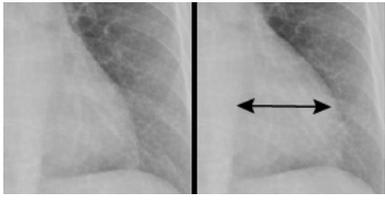 | 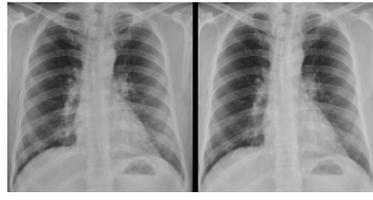 | 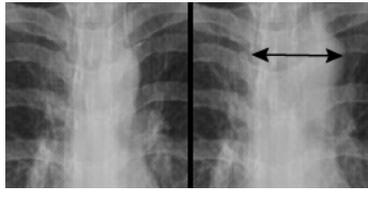 |
|---|---|---|---|
| Prediction Task | Normal / Abnormal | Normal / Abnormal | Normal / Abnormal |
| Attribute location ("Where")" | Left ventricle | Exposure | Upper mediastinum |
| Human description of the attribute ("What")" | Left ventricular enlargement, demonstrated by leftward displacement of the left heart border, with increasing CP of abnormal CXR. | More apparent over exposure images (darker images) with increased likelihood of abnormal CXR. | Mediastinal widening or prominent aortic knob with increasing CP of abnormal CXR. |
| Consolidated Panel Notes | Left ventricular enlargement often occurs in the setting of congestive heart failure, ischemic heart disease, or hypertension. Not likely to be regular beating of heart because enlargement is focal. There are severe racial disparities in heart disease and hypertension conditions, so would want to think about this in the context of the demographic makeup of the data. | Portable (anteroposterior) CXRs are usually obtained at the bedside with a patient in a supine position (usually taken in-patient, when people are more sick, less mobile, higher acuity). When compared to standard posteroanterior CXRs obtained with a standing patient in a radiology room, portable CXRs tend to vary more in quality (more likely to be over exposed, although can be under or over exposed). The association may be affected by which patient types are likely to get CXRs in a hospital setting bedside vs radiology room, and which healthcare providers/technicians may be administering the CXRs. Other considerations include whether there was a change in the way photos were taken over time due to changing hospital protocols (e.g., COVID-19 special protocols) and that digital radiographs tend to be more overexposed than (older) film screen images. | Atherosclerosis and older age are associated with elevated disease risk. Although mediastinal width can sometimes be exaggerated by patient positioning or portable imaging, an enlarged or tortuous thoracic aorta is a common cause of true mediastinal widening or prominent aortic knob in elderly individuals, particularly those with atherosclerosis. It is possible that this association be an artifact of the age range of the participants (i.e. average age was >50). Positioning or Portable imaging is not likely the issue because the heart does not expand along with the mediastinal widening. |
| Areas of Exploration from Interdisciplinary Expert Panel | Would association with CP still exist if model were run on a dataset where the distribution of disease was different? | Would we see the same association with the CP if we looked at both sitting and standing X-ray separately? Do we find that clinical sites, in particular geographies, distribution of | Would association with CP still be present if model were run on datasets with a different distribution of age (e.g., lower average age) |

| | | patient SES, or physician/technician type are also associated with exposure and could represent confounding? | |
|---|---|---|---|

## Supplementary Table 2c

### Fundus photo attributes

| | | | |
|---|---|---|---|
| Images of increased / decreased attribute magnitude | 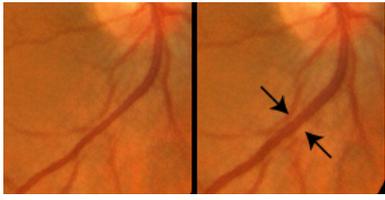 | 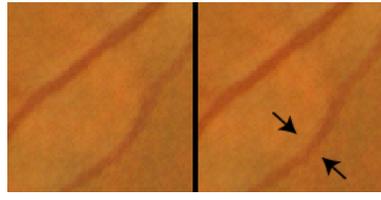 | 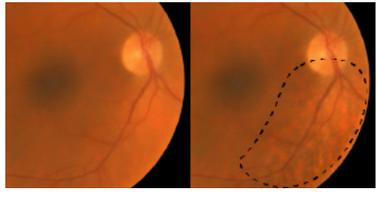 |
| Prediction Task | Smoking Status (yes/no) | Systolic BP | Sex |
| Attribute location ("Where")" | Retinal vessels | Retinal arteries | Choroid |
| Human description of the attribute ("What")" | Retinal vein dilation with increasing CP of being a smoker | Arteriolar narrowing with increasing CP of elevated systolic BP | Increase in nasal, temporal, and inferior choroidal vasculature visibility with increasing CP of male gender |
| Consolidated Panel Notes | Retinal venular caliber is associated with cardiovascular disease, which is also associated with smoking. Smoking, cardiovascular disease, and diabetes are so intertwined, it would be difficult to disentangle. For instance, cardiovascular disease is strongly associated with smoking, but it would be difficult to find a dataset with smokers who don't have cardiovascular disease. Even if such datasets exist, social, cultural, and environmental factors between participants in those datasets would likely vary greatly, creating additional issues of confounding. | StylEx visualizes this ubiquitous sign of hypertension. Retinal artery narrowing is a manifestation of the vasoconstrictive phase of early systemic hypertension. | Investigate further to improve the model or to understand why the model can make this prediction. This attribute associates greater choroidal vasculature visibility with increased probability of male sex, which is the converse of what previous research in this area has suggested. There may be differences in the dataset, related to distribution of myopia and fundus pigmentation within male and female populations, which may drive the differences identified by the model. Sex is likely serving as a proxy for factors unrepresented in the dataset. For instance, dataset has unknown distribution of factors associated with sex, skin tone, myopia, and fundus pigmentation, making it difficult to determine what model is picking up on. The association is not likely related to presence of autoimmune disease or eye irritants. |
| Areas of Exploration from Interdisciplinary Expert Panel | Would this association with the CP still be present if it were possible to isolate the effect of smoking, distinct from factors such as cardiovascular disease, diabetes, and socioeconomic status? Would this association be present if model were run on datasets with varying | | Would association with CP still exist if the classifier were run on a dataset with greater variation in sex, skin tone, myopia, or fundus pigmentation? |

|  | severity of cardiovascular disease or diabetes? |  |  |

## Supplementary Table 2d

### External eye photo attributes

| Images of increased / decreased attribute magnitude | 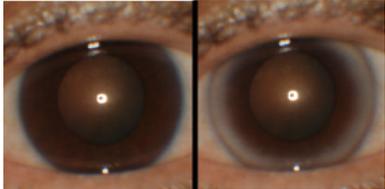 | 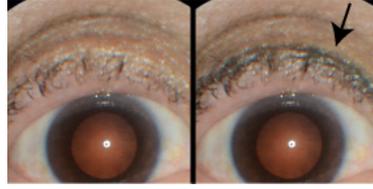 | 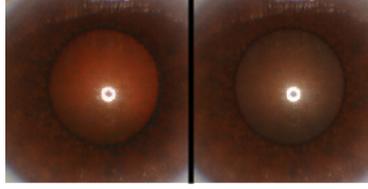 |
|---|---|---|---|
| Prediction Task | HbA1c >= 9 | Hgb | Cataract |
| Attribute location ("Where")" | Peripheral cornea | Eyelid | Pupil |
| Human description of the attribute ("What")" | Increase in corneal arcus thickness from the limbus with decreasing probability of elevated HA1c | Increasing upper eyelid eyeliner thickness and density with increasing CP of low Hgb | Dimmer red reflex with higher CP of cataract |
| Consolidated Panel Notes | Selection bias is likely present in this statistical association, most apparently related to clinic type represented in the datasets (e.g., survivorship bias- people with more severe diabetes have been referred to other practitioners; socioeconomic status of patients is skewed towards lower socioeconomic status, given that these are clinics are safety-net clinics). This association is likely not due to variations in eye movement and camera exposure. | The eyeliner attribute identifies a known cultural association between makeup usage and low hemoglobin levels with the same demographic factors of sex and age. Additionally, the dataset was not racially and ethnically diverse, warranting questions of whether this same association would have been observed with darker skin tones or with use of different colors of eyeliner. | Lens opacification (cataract) obscures the red-orange reflection of the retina (red reflex) with ophthalmoscopy or photography. Thus, photographs of cataracts exhibit a reduced red reflex as compared to images without cataract. The association is not likely due to affect of camera exposure or ambient light due to consistent camera and protocol. |
| Areas of Exploration from Interdisciplinary Expert Panel | Would association still be present if model were run on datasets with greater distribution of disease severity or socioeconomic status? | Would association with the CP still be present if model were run on datasets that were subset by a rough proxy for eye makeup usage (i.e., sex) or if model had greater distribution images with varying levels of contrast (e.g., dataset including varying skin tones using dark eye liner or dataset including people of similar skin tones using metallic or lighter eyeliner colors)? | Would association with the CP still be present if model were run on datasets with greater distribution of disease severity? |

## Supplementary Table 2e
External eye photo attributes

| Images of increased / decreased attribute magnitude | 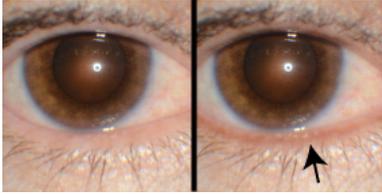 | 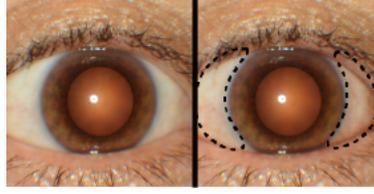 | 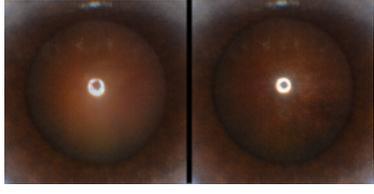 |
|---|---|---|---|
| Prediction Task | HbA1c >= 9 | Hgb | Cataract |
| Attribute location ("Where")" | Eyelid margin | Conjunctiva | Lens |
| Human description of the attribute ("What")" | Increasing eyelid margin pallor with increasing probability of elevated HA1c | Decreased conjunctival vessel prominence with increasing CP of low Hgb | Development and extension of cortical cataract spokes from peripheral to central lens with increasing probability of cataract |
| Consolidated Panel Notes | The eyelid margin attribute may be depicting pathophysiologic signs of meibomian gland disease (MGD), which is associated with increasing chance of diabetes. Signs and symptoms of MGD or dysfunction are noted with increased frequency in individuals with diabetes. Manifestations of MGD include gland dropout, atrophy, inspissation, and migration of the marx line. There are several other potential confounders such as environmental factors and medications that should be considered, such as toxic exposures to medication and irritants (see here). Age and sex may be proxies for social and cultural practices related to hygiene, use of products such as make up and other topicals. | The biologic explanation is consistent with the StylEx attribute that associates reduced conjunctival vessel visibility with anemia. Conjunctival pallor is a well known sign of anemia and is attributed to a reduction of circulating oxyhemoglobin in tissue vasculature. The association is not likely due to affect of camera exposure or ambient light due to consistent camera and protocol. | Lens opacification (cataract) obscures the red-orange reflection of the retina (red reflex) with ophthalmoscopy or photography. Thus, photographs of cataracts exhibit a reduced red reflex as compared to images without cataract. The association is not likely due to affect of camera exposure or ambient light due to consistent camera and protocol. |
| Areas of Exploration from Interdisciplinary Expert Panel | Would association still be present if model were run on datasets that were subset by age and by sex? Which specific cultural, social, and behavioral practices for which age and sex may be representing a proxy for in this dataset? | | Would association with the CP still be present if model were run on datasets with greater distribution of disease severity? |

# Supplementary Table 3: Panelist backgrounds

| Domain | Panelists |
|---|---|
| CXR | <ul><li>Radiologist, 18 YOE</li><li>ML Engineer 1, 17 YOE</li><li>ML Engineer 2, 13 YOE</li><li>Socio-technical expert, epidemiologist, health services researcher, 8 YOE</li><li>Social-behavioral epidemiologist, informaticist, 18 YOE</li><li>HCI expert, 14 YOE</li><li>Scientist with experience across medical specialties and modalities, 16 YOE</li></ul> |
| Fundus & External eye | <ul><li>Ophthalmologist, 9 YOE</li><li>Ophthalmologist, 12 YOE</li><li>ML Engineer 1, 17 YOE</li><li>ML Engineer 2, 13 YOE</li><li>Socio-technical expert, epidemiologist, health services researcher, 8 YOE</li><li>Social-behavioral epidemiologist, informaticist, 18 YOE</li><li>HCI expert, 14 YOE</li><li>Scientist with experience across medical specialties and modalities, 16 YOE</li></ul> |

## Supplementary Table 4: Numbers of attributes selected for the expert panel

The number of attributes selected for discussion of the expert panel out of the total attributes which passed the automated filtering (Methods). Due to time constraints we considered only a smaller set for each classifier (see Supplementary Note 4).

| Image modality | Classifier task | No. of attributes passed by automatic filtering | No. of attributes evaluated in detail by the expert panel |
| --- | --- | --- | --- |
| External Eye | HbA1c | 3 | 2 |
| | Hgb | 5 | 2 |
| | Cataract | 4 | 2 |
| Fundus | Systolic BP | 4 | 1 |
| | Smoking Status (yes/no) | 3 | 1 |
| | Sex | 6 | 1 |
| CXR | Normal / Abnormal | 8 | 3 |
| | Race | 9 | 3 |